\documentclass[11pt,twoside]{article}
\usepackage{newpasp,epsf}
\def\edcomment#1{\iffalse\marginpar{\raggedright\sl#1\/}\else\relax\fi}
\reversemarginpar

\begin{document}
\title{A high-resolution multi-wavelength study of the jet in 3C\,273}
\author{Sebastian Jester, Hermann-Josef R\"oser and Klaus
Meisenheimer}
\affil{Max-Planck-Institut f\"ur Astronomie, K\"onigstuhl 17,
69117 Heidelberg, Germany}
\author{Rick Perley}
\affil{NRAO, Socorro, NM 87801, USA}
\author{Simon Garrington}
\affil{Jodrell Bank Observatory, Macclesfield, Cheshire SK11 9DL, UK}

\begin{abstract}
We present HST images at 622\,nm and 300\,nm of the jet in 3C\,273 and
determine the run of the optical spectral index at 0\farcs2 along the
jet.  We find no evidence for localized acceleration or loss sites,
and support for a little-changing spectral shape throughout the jet.
We consider this further evidence in favour of a distributed
acceleration process.
\end{abstract}

Of the plethora of known extragalactic radio jets, optical emission
has to date been observed from only about 15 extragalactic jets.  The
radio and optical emission observed from terminal hot spots of radio
jets can be explained by first-order Fermi acceleration at a strong
shock in the jet (the bow shock) (Meisenheimer \& Heavens 1986;
Heavens \& Meisenheimer 1987; Meisenheimer et al. 1989; Meisenheimer,
Yates, \&R\"oser 1997).  But it is not clear that the optical
synchrotron emission from the jet \emph{body}, extending over tens of
kiloparsecs in some cases, can equally be explained by acceleration at
strong shocks inside the jet.  Electrons with the highly relativistic
energies required for the emission of optical and UV synchrotron
radiation have a very short lifetime which is much less than the
light-travel time down the jet body in 3C\,273, for example.  We have
embarked on a detailed study of the jet in 3C\,273 using broad-band
observations at various wavelengths obtained with today's best
observatories in terms of resolution: the VLA (with a planned
combination with MERLIN data) and the HST.  Here, we present optical
and near-UV images obtained with HST, and the run of the optical
spectral index along the jet of 3C\,273.

The jet was observed with WFPC2 in March and June 1995, using the
Planetary Camera (PC) with a pixel size of 0\farcs0455 (proposal ID
5980).  The total exposure time was 35\,500\,s through filter F300W
(ultra-violet, centered at 300\,nm) and 10\,000\,s through F622W (red
light, 620\,nm). Figure 1 shows the reduced, background-subtracted sum
images.
\begin{figure}
\plotone{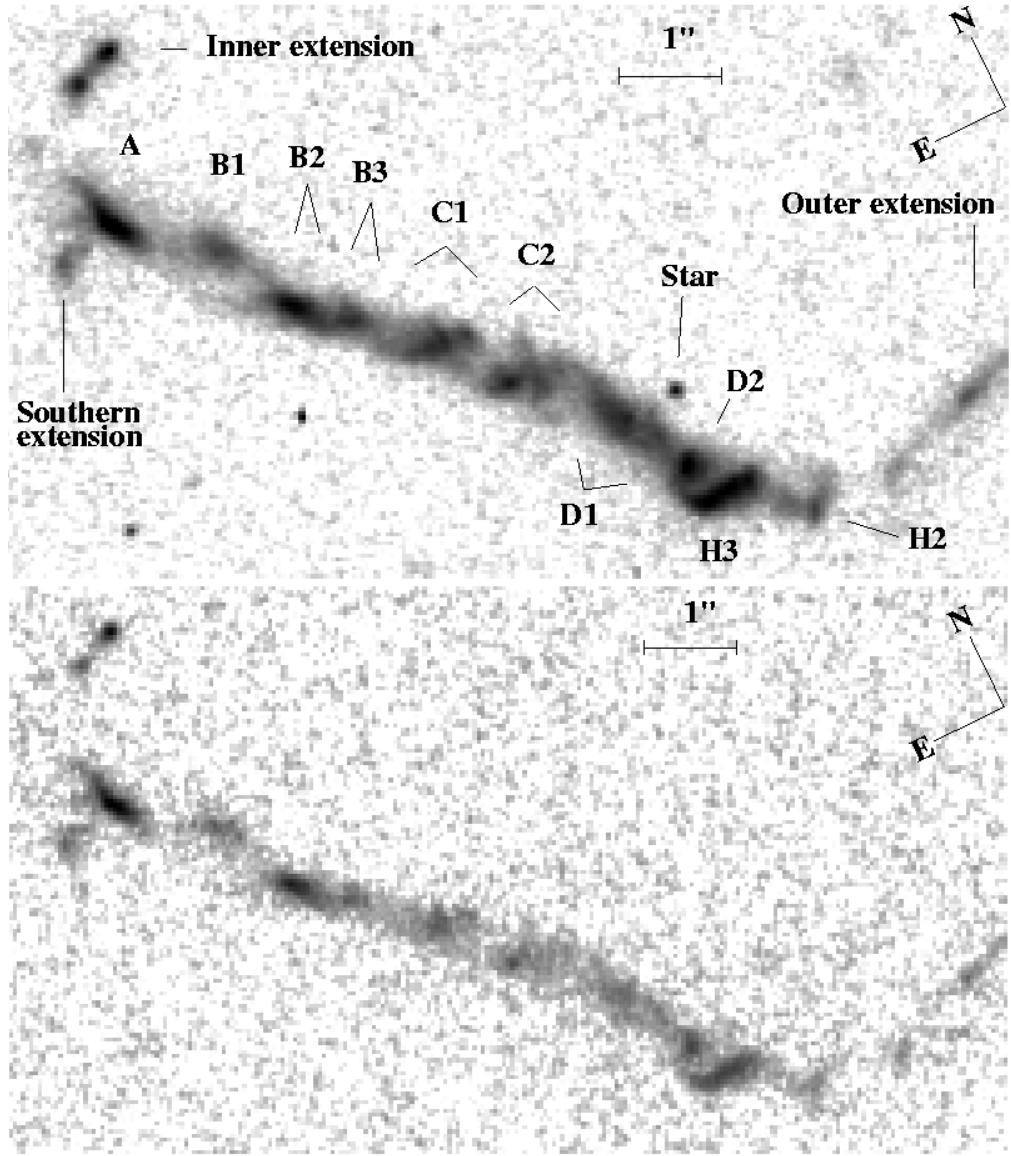}
\caption{The optically visible part of the jet in 3C\,273 with WFPC2
through filters F622W (red light, top) and F300W (UV, bottom)
after background subtraction. Logarithmic grey-levels run from 0 to
0.04$\mu$Jy/pixel (red) and 0 to 0.014$\mu$Jy/pixel (UV),
respectively.  The labelling of the jet knots as introduced by
Leli\`{e}vre et al.\ (1984) and extended by R\"oser \& Meisenheimer
(1991), together with the hot spot nomenclature from Flatters \&
Conway (1985), is also shown.  H2 is the location of the radio hot
spot.  The core is 10\arcsec\ towards the Northeast from knot A.}
\end{figure}

Independently of whether a spectrum actually does follow a power law
$F_{\nu} \propto \nu^{\alpha}$ over any range of frequencies, a
two-point spectral index can be defined between any two surface
brightness measurements as
\begin{math}
\alpha_{12} = \ln\frac{S_1}{S_2}/\ln\frac{\nu_1}{\nu_2}.
\end{math} 
We calculated a map of the optical spectral index $\alpha_{RU}$
between red and ultraviolet at 0\farcs2 beam size from Fig.~1.
(Details of the data reduction and the alignment process used to avoid
systematic errors in the spectral index will be given in our
forthcoming paper (Jester et al.\ 2000, \emph{in prep.}).)  From the
jet image (Fig. 1), we extracted a trace of the brightness profile by
locating the brightness maximum in slices perpendicular to the jet's
mean position angle of 222\fdg2. Figure 2 shows this profile and the
run of the optical spectral index at the brightness maximum.
\begin{figure}
\plotone{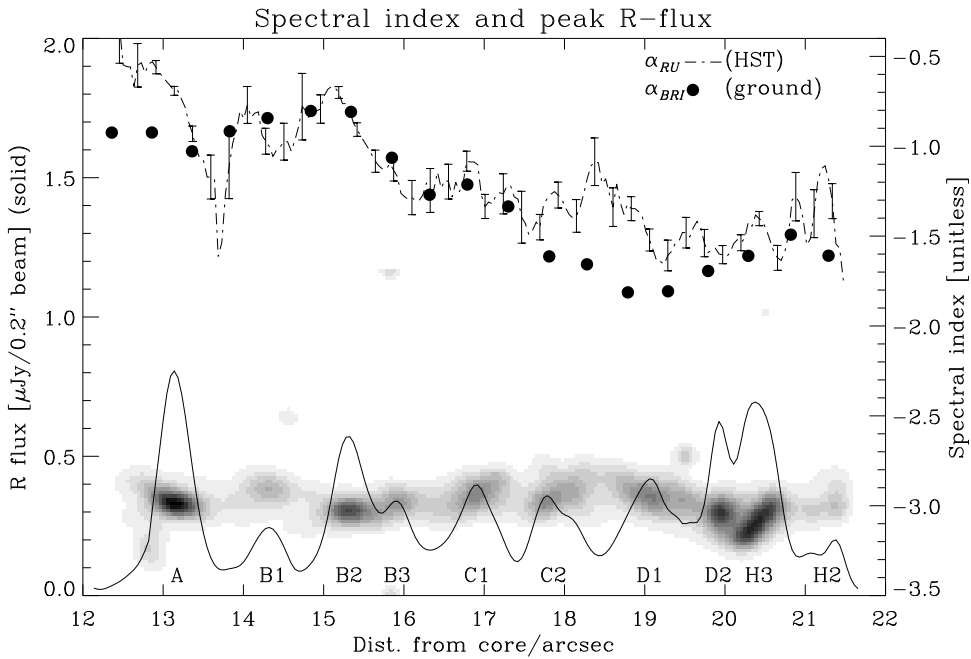}
\caption{Run of the $R$-band brightness (622\,nm) and optical spectral
index along the outer half of the jet in 3C\,273, for a 0\farcs2
beam. $\alpha_{RU}$ was determined from Fig.~1, while $\alpha_{BRI}$
for 1\farcs3 beam size is taken from R\"oser \& Meisenheimer (1991).}
\end{figure}

The spectral index variations are very smooth over the entire jet at
our resolution of 0\farcs2 ($640 h^{-1}_{60}$\,pc).  There is no
abrupt change of $\alpha_{RU}$ inside the bright regions.  Our
determination of the spectral index between 622\,nm and 300\,nm shows
a good overall agreement with the spectral index $\alpha_{BRI}$
obtained from measurements at 450\,nm, 650\,nm and 860\,nm (from
R\"oser \& Meisenheimer 1991).  The overall trend is a steepening of
the spectral index, from $-0.5$ at the inner end of the jet (A) down
to $-1.5$ at D2/H3 just before the radio hot spot.  This corresponds
qualitatively to the outward decrease of the cutoff frequency as
determined from synchrotron spectra fitted to the brightest jet knots
at 1\farcs3 resolution (Meisenheimer, Neumann, \& R\"oser 1996).  The
only discrepancy between $\alpha_{RU}$ and $\alpha_{BRI}$ occurs in A
and C2/D1. Clearly, flux from neighbouring knots contributes to each
point at 1\farcs3 resolution.  There may be further
reasons, \emph{e.\,g.}, the fact that the jet widens at
longer wavelengths: there is extended emission around the jet in the
infrared (Neumann, Meisenheimer, \& R\"oser 1997), and a larger beam
will pick up some flux from this steeper component also at 622\,nm.

The smooth spatial changes and the similarity of the optical spectral
index at vastly different scales imply a little-changing spectral
shape over the entire jet.  We are unable to find localized
acceleration centres and conclude that there is a smooth distribution
of accelerated particles.

A full, quantitative discussion of the optical spectral index will be
given in future publications.  There, we will examine in detail the
shape of the synchrotron spectrum from radio to UV frequencies by
fitting spatially resolved spectra according to Meisenheimer et
al. (1989).  This will enable the detection of deviations of the
spectral shape from a simple power law with cutoff.  For example, a UV
flux point above a synchrotron cutoff fitted to the
radio-to-optical data will be a hint for a second high-energy electron
population producing the X-ray emission (R\"oser et
al. 2000).  The mapping of the maximum particle energy determined from
the spectra will further constrain the locality of the acceleration
process.

\acknowledgements These results are based on observations made with
the NASA/ESA Hubble Space Telescope, obtained at the Space Telescope
Science Institute, which is operated by the Association of
Universities for Research in Astronomy, Inc.\ under NASA contract
No.~NAS5-26555.

\end{document}